\documentclass [11pt]{article}
\usepackage{epsfig}

\begin{document}
 \def\la{\langle}
 \def\ra{\rangle}
 \def\gtt{\overline{t}t\gamma}
 \def\ztt{\overline{t}tZ}
\hfill {WM-06-105}

\vskip 1 in \baselineskip 24pt

{\Large
   \bigskip
   \centerline{Top Pair Production in Randall-Sundrum Models}
 }
\def\bar{\overline}

\centerline{Erin De Pree\footnote{Email: ekdepr@wm.edu} and Marc
Sher\footnote{Email: sher@physics.wm.edu}}
\centerline {\it Particle Theory Group}
\centerline {\it Department of Physics}
\centerline {\it College of William and Mary, Williamsburg, VA 23187,
USA}
\bigskip

{\narrower\narrower If the lowest
lying Kaluza-Klein states in Randall-Sundrum (RS1) models have masses 
in the $10-100$ TeV range,
direct production of these states at the LHC or ILC is impossible,
and electroweak precision measurements may not be sufficiently
sensitive.   We address the possibility that high-precision
measurements of top pair production at the ILC may provide the first 
evidence
of these states.   We consider RS1 models with fermions on and off the
brane, with  bulk left and right handed mass terms, discuss
brane kinetic terms and calculate 
corrections
to top pair production in these models.}

\vskip 1.0cm
\newpage
\section{Introduction}

For the past quarter of a century, two of the most promising 
solutions to
the gauge hierarchy problem have been supersymmetry and
technicolor.   These extensions of the standard model have provided
some of the primary motivations for the LHC and the ILC, and have
provided a rich framework for studying beyond-the-standard-model
phenomenology.

An alternative approach was provided several years ago by the
Randall-Sundrum (RS1) model \cite{randallsundrum}.   In this model,
spacetime is five-dimensional, with one dimension compactified on an
$S_{1}/Z_{2}$ orbifold.  The five-dimensional bulk geometry is a slice
of anti-de Sitter ($AdS_{5}$) space.   At the fixed points of the
orbifold (at $y=0,\pi R$), the slice is bounded by 3-branes of equal
and opposite tension.   The brane at $y=0$ is referred to as the
Planck brane, while the brane at $y=\pi R$ is referred to as the TeV
brane.  The curvature scale, $k$, and the length of the $AdS_{5}$
slice, $\pi R$, are expected to be of the order of the Planck mass,
$M_{P}$ and
its inverse, respectively.   The geometry then induces a effective scale on the TeV
brane of the order of $M_{P}e^{-\pi kR}$.   For $kR\simeq 11$, which
is not particularly ``fine-tuned'', this scale is of the order of a
TeV.    If the
Higgs field(s) live on the TeV brane, then the electroweak scale is
naturally generated.  Thus, the hierarchy problem is solved.   Several
very nice reviews of the model, as well as many of the issues
discussed in the rest of this section, can be found in Ref.
\cite{reviews}

In the original model, only gravity propagated in the bulk and the
standard model fields were confined to the TeV brane.  Nonetheless,
this leads to interesting collider effects from Kaluza-Klein (KK) 
graviton
exchange \cite{graviton}.  It was realized at an early stage that a
much richer phenomenology would arise if one allowed some of the
standard model fields to propagate in the bulk.

Initially, the effects of gauge bosons in the bulk (with the Higgs
field and fermions still confined to the TeV brane) were
considered \cite{pomarol,dhr}.   In this model, the couplings of the
fermions to the KK excitations of the gauge bosons are enhanced
relative to the couplings to the zero-mode gauge bosons by a factor
of $\sqrt{2\pi kR}\simeq 8.4$.   These large couplings cause serious
constraints \cite{chang,hub1,erlich} from precision electroweak 
measurements, with
bounds ranging from $10-25$ TeV on the mass of the lowest lying
KK excitation of the gauge bosons.  Such a high mass would be beyond
the reach of the LHC, and would also reintroduce the hierarchy
problem (although at a much smaller level of fine-tuning).

One method of relaxing these constraints, with fermions still on
the TeV brane, is to include brane-localized kinetic terms for the
gauge fields.  These terms should be present in
general \cite{georgi}.  Their effects on couplings and masses were
shown to be substantial in flat space \cite{ctw}, and an
analysis \cite{dhr2} in
the RS model showed that the lower bound on the lightest KK 
excitation mass could be
substantially smaller.

An alternative approach to relaxing the constraints is to allow
fermions to propagate in the bulk.  This also gives the exciting
possibility of explaining the large fermion mass hierarchies.   With
fermions in the bulk, the bounds from electroweak precision data were
somewhat ameliorated \cite{hub1,gher,dool,dhr3,hub2,care,care2}.   In addition,
since fermions are in the bulk, the couplings of the fermions to the
Higgs boson (which remains on the TeV brane) can be substantially
suppressed by the geometric warp factor 
\cite{gher,gross,hub3,hub4}.   For fermions near the TeV
brane, the suppression is small, but for fermions far from the TeV
brane, the suppression can be exponential, leading to large fermion
mass hierarchies.  The observed fermion mass hierarchy then becomes a
matter of fermion geography.  Huber \cite{hub4} has shown explicitly 
how simple
parameters of $O(1)$ can lead to the observed fermion mass hierarchy
and mixings.

As shown by Agashe, et al. \cite{agashe1}, the model still had large
contributions  to the T parameter in electroweak
radiative corrections, forcing the KK scale to still be out of reach
of the LHC.  It also had large contributions to $Z\rightarrow 
\bar{b}b$.
The reason is that the large top quark mass forces the top quark to
be near the TeV brane, so that it can interact strongly with the
Higgs.  But since the left-handed top is paired with the left-handed
bottom, the left-handed bottom will have to be near the TeV brane, and
that leads to larger corrections to the $Z\rightarrow \bar{b}b$ rate.
They showed that imposing a custodial isospin symmetry in the bulk (by
enlarging the gauge group to $SU(2)_{L}\times SU(2)_{R}\times
U(1)_{B-L}$) solves both of these problems, and allows the lowest
lying KK states to have masses as low as a few TeV, within range of
the LHC.   These models are attractive in that the custodial isospin
gauge symmetry of the bulk can be dual, through the AdS/CFT
correspondence, to a global isospin symmetry of the CFT.

There are other alternatives.   Hewett, Petriello and Rizzo \cite{hpr}
consider putting the first two families in the bulk and the third on
the brane, and alleviate these problems.  This paper was the first to 
consider top pair production in Randall-Sundrum models at a linear 
collider, although it was in the context of the model with the third 
generation on the brane and used a  common mass parameter for the 
other fermions.  More recently, Carena et
al. \cite{care2} show that brane kinetic terms for the fermions can also
give good fits for relatively light KK masses.    An
introduction to brane kinetic terms can be found in Ref.
 \cite{delaguila}.    A summary of many of these issues, including 
flavor changing neutral currents, can be found in Ref. \cite{moreau}, 
where it is pointed out that the KK mass scale could be lowered to the 
few
TeV mass scale without problems with precision electroweak data.

Our approach in this paper is somewhat different.  We will not
attempt to find ways to lower the KK masses to the range of the LHC,
but will consider the possibility that these masses are in the
$10-100$ TeV range.   In this case, they will be out of reach of the
LHC and ILC, and (except possibly in the lower end of the range for
some models) will be insensitive to electroweak precision 
measurements (and any sensitivity can be eliminated with one of the 
techniques discussed above).
Of course, there will be a hierarchy problem, although substantially
less of a problem than in standard grand unified theories, and we
will not address that issue.   In this scenario, what would the first
experimental evidence be?  Since the top quark is close to the TeV
brane, effects of KK states on top pair production would be the most
pronounced, and thus could be the first signature (more likely at the
ILC, where higher precision measurements can be made).  In this work, 
we
study top pair production in a variety of RS models, and determine the
reach of KK masses expected at the ILC.

In Section 2, the RS models are presented. 
In Sections 3, we consider only the effects of KK gauge bosons,
ignoring KK fermions.  In Section 4, the effects of KK fermions and 
of brane kinetic terms are considered.  Finally Section 5 contains our
conclusions.

\section{The Models}

The metric of the Randall-Sundrum model \cite{randallsundrum} is given
by
\begin{equation}
    ds^{2}= e^{-2\sigma(y)}\eta_{\mu\nu}dx^{\mu}dx^{\nu} - dy^{2}
    \end{equation}
    where $\sigma(y)=k|y|$, $k$ is related to the curvature of the
    AdS space, $\eta_{\mu\nu}$ is the flat-space metric,
and $y$ is the fifth coordinate.  The fifth dimension is compactified
on an $S^{1}/Z_{2}$ orbifold bounded by branes at the fixed points
$y=0$ and $y=\pi R$.   In this section, we present the masses and
couplings of gauge bosons and fermions, when they propagate in the
bulk.  More detailed derivations of these results can be found in
references cited in the last section.

The equation of motion for a bulk gauge field is given
by \cite{pomarol,dhr,hub1,hub2,hub4}
\begin{equation}
    {1\over\sqrt{-G}}\partial_{M}(\sqrt{-G}\ G^{MN}G^{RS}F_{NS})
    -M^{2}_{A}G^{RS}A_{S}=0
\end{equation}
where $M_{A}$ arises from spontaneous symmetry breaking, $G^{MN}$ is
the above metric and $\sqrt{-G}=e^{-2\sigma}$.  This can be
rewritten as
\begin{equation}
[e^{2\sigma}\eta_{\rho\nu}\partial^{\rho}\partial^{\nu}+e^{2\sigma}\partial_{5}(e^{-2\sigma}
\partial_{5})-M_{A}^{2}] A(x_{\mu},y) = 0
\end{equation}
The Higgs field is localized on the TeV brane, and thus
$M^{2}_{A}={1\over 2}g_{5}^{2}v^{2}\delta(y-\pi R)$.  The vacuum
expectation value is of the order of the Planck mass.

Decomposing the gauge field (using the gauge
$A_{5}=\partial_{\mu}A^{\mu}=0$), one has
\begin{equation}
A(x_{\mu},y)={1\over \sqrt{2\pi
R}}\sum_{n=0}^{\infty}A^{(n)}(x_{\mu})f_{n}^{A}(y)
\end{equation}
where the orthogonality condition is
\begin{equation}
    {1\over 2\pi R}\int_{-\pi R}^{\pi R}dy 
f_{n}^{A}(y)f_{m}^{A}(y)=\delta_{mn}
    \end{equation}

Plugging the decomposition into the equation of motion, one can solve
the equation and find \cite{pomarol,dhr,hub4}
\begin{equation}
    f^{A}_{n}(y) = {e^{\sigma}\over N}\left[ J_{1}\left({m_{n}\over
    k}e^{\sigma}\right) + b_{1}(m_{n}) Y_{1}\left({m_{n}\over k}\right)\right]
    \end{equation}
The values of $m_{n}$ and $b$ are given by the boundary conditions,
and $N$ by the normalization condition.
Note that the mass term does not enter into this equation; it will
only affect the boundary conditions at the TeV brane.   Imposing
these conditions gives the zero-mode mass \cite{hub4}
\begin{equation}
    m^{2}_{0}= g^{2}_{5}v^{2}e^{-2\pi kR}\left(1+{\cal
    O}(g_{5}^{2}v^{2}e^{-2\pi kR}/M^{2}_{1})\right)
\end{equation}
where $M_{1}$ is the KK scale.  Note that a gauge hierarchy naturally
appears.  The higher order correction causes a tree-level shift in
the W and Z masses, affecting electroweak precision data if the KK
scale is too small, leading to many of the bounds noted in the
previous section.  The masses of the KK-excitations of the gauge
bosons are related to zeroes of the Bessel functions.  One
can add brane kinetic terms for the gauge bosons, as will be
discussed in Section 4.

If the fermions are on the TeV brane, then, as shown in Refs.
\cite{pomarol,dhr}, their couplings to the gauge bosons are of the 
form
\begin{equation}
    {\cal L}=-g\overline{\psi}\gamma^{\mu}\left(
    A^{(0)}_{\mu}+\sqrt{2\pi
    kR}\sum_{n=1}^{\infty}A_{\mu}^{(n)}\right)\psi
    \end{equation}
which gives an enhancement of $\sqrt{2\pi kR}\simeq 8.4$ in the
coupling.   This changes substantially if the fermions are in the 
bulk.

When fermions are in the bulk \cite{gher,gross,hub4}, they can have 
two possible
transformation properties under the orbifold $Z_{2}$ symmetry:
$\psi=\pm \gamma_{5}\psi$.  As a result, $\overline{\psi}\psi$ is
odd under the $Z_{2}$, and thus the Dirac mass term must originate
from coupling to a $Z_{2}$ odd scalar field.  This mass term can then
be written as $m_{\psi}=c{d\sigma\over dy}$, where $\sigma=k|y|$.  As
we will see shortly, the parameter $c$ will be crucial in determining
the properties of the fermions.

As
before, one can expand the fields and determine the wavefunctions and
masses of the fermions.   One expands
\begin{equation}
    \psi(x^{\mu},y)={1\over 2\pi
    R}\sum_{n=0}^{\infty}\psi^{n}(x^{\mu})e^{2\sigma}f_{n}(y)
    \end{equation}
where the normalization condition is
\begin{equation}
    {1\over 2\pi R}\int_{-\pi R}^{\pi R}dy\
    e^{\sigma}f_{m}(y)f_{n}(y) = \delta_{mn}
    \end{equation}
and the factor of $e^{2\sigma}$ comes from the spin
connection.

Plugging into the Dirac equation, one finds the zero mode wave
function is simply (we suppress flavor indices and neglect flavor 
mixing)
\begin{equation}
    f_{0}(y) = {e^{-c\sigma}\over N_{0}}
    \end{equation}
and the KK-fermion wave functions are
\begin{equation}
    f_{n}(y) = {e^{\sigma/2}\over
    N_{n}}\left[ J_{\alpha}\left({m_{n}\over
    k}e^{\sigma}\right) + b_{\alpha}(m_{n}) Y_{\alpha}\left({m_{n}\over k}\right)\right]
    \end{equation}
where $\alpha=|c\pm {1\over 2}|$ for $\psi_{L,R}$.   The masses and
$b_{\alpha}$ are given by the boundary conditions.

The zero-mode wave function is sufficiently simple that the
normalization constant $N_{o}$ can be determined easily to be
\begin{equation}
    N^{2}_{o}={e^{2\pi kR(1/2-c)}-1 \over 2\pi kR(1/2-c)}
    \end{equation}
From this, one can see that if $c>1/2$, the zero mode fermions will be
localized near the Planck ($y=0$) brane, while for $c<1/2$, they will
be localized near the TeV ($y=\pi R$) brane.

The zero modes acquire mass through coupling to the Higgs field on
the TeV brane (here, we include flavor indices)
\begin{equation}
    m_{ij}=\int_{-\pi R}^{\pi R}{dy\over 2\pi
   R}\lambda_{ij}^{5}\langle H(y)\rangle f_{0iL}(y)f_{0jR}(y)
   \end{equation}
and using $\langle H(y)\rangle = v\delta(y-\pi R)/k$, one finds
\begin{equation}
    m_{ij}={\lambda_{ij}^{4}v\over \pi kR}f_{0iL}(\pi R)f_{0jR}(\pi R)
    \end{equation}
where the dimensionless 4-D coupling
$\lambda_{ij}^{4}=\lambda_{ij}^{5}\sqrt{k}$.

This demonstrates how a huge fermion mass hierarchy can arise.
For $c<1/2$, the wave function $f_{0}(\pi R)$ varies as
$\sqrt{1-2c}$, but for $c>1/2$ varies as $e^{-c\pi kR}$.
Since $\pi kR\simeq 35$, this exponential suppression can lead to a
hierarchy.  Huber \cite{hub4} shows explicitly how mild variations in
$c$ can lead to the observed mass spectrum, and can also lead to
reasonable flavor mixing.

The couplings between gauge bosons and fermions come from the 5-D term
\begin{equation}
    \int d^{4}x dy
    \sqrt{-G}g_{5}\overline{\psi}(x,y)i\gamma^{\mu}A_{\mu}(x,y)
    \psi(x,y)
\end{equation}
which induces 4D-couplings
\begin{equation}
    g_{ijn}={g_{5}\over (2\pi R)^{3/2}}\int_{-\pi R}^{\pi
    R}e^{\sigma}f_{i}(y)f_{j}(y)f_{n}^{A}(y)dy
\end{equation}
From this, we can now determine all gauge-boson couplings to fermions.

Note that for a zero-mode massless gauge boson, $f^{A}_{0}=1$, and the
result just gives the normalization condition, giving
$g_{ij0}=\delta_{ij}g_{5}/\sqrt{2\pi R}$, thus fermion couplings to
the zero-mode are KK-level conserving.

For our calculation, we will need the coupling of a KK-gauge boson to
zero-mode fermions, which is then\footnote{In Ref. \cite{gher}, the
first factor of $e^{\sigma}$ in the integral is missing--this is
entirely typographical and does not affect their results.}
\begin{equation}
    g^{(n)}=g\left( {1-2c\over e^{(1-2c)\pi kR}-1}\right){k\over
    N_{0}}\int_{0}^{\pi R}dy\
    e^{\sigma}e^{(1-2c)\sigma}\left[ J_{1}\left({m_{n}\over
    k}e^{\sigma}\right) + b_{1}(m_{n}) Y_{1}\left({m_{n}\over k}\right)\right].  
    \label{g(n) coupling}
\end{equation}
These are plotted in Ref. \cite{gher} for $n=1,2,3$ as a function of
$c$.  For $c$ large and negative (so the fermion is very close to the 
TeV
brane), the coupling ratio reaches $\sqrt{2\pi kR} \simeq 8.4$, as
discussed earlier.  As $c$ increases, they become smaller, vanishing
in the conformal limit $c=1/2$, and then reach a constant value of
approximately $-0.2$ for $c>1/2$.

This scenario is very attractive, due to the manner in which the
fermion mass hierarchy naturally arises.  We can see that fermions
near the TeV brane couple more strongly than those away from the
TeV brane.  Since the top quark is closest to the TeV brane, one
expects the biggest effects to arise in top-quark processes, and if
the KK-scale is much larger than 10 TeV, these processes may be the
first signature.

We now turn to top pair production, and first consider only the
effect of KK-gauge bosons.  Note that in the absence of
brane kinetic terms, the masses
of the KK-fermions (for a given value of the Dirac mass term)
are related to those of KK-gauge bosons (through
zeroes of Bessel functions), and such a
consideration is not realistic.  But since brane kinetic
terms can decouple the masses, such a separation is consistent.  
Following
the discussion of the effects of KK-gauge bosons, we will turn to
those of
KK-fermions.

\section{Effects of KK Gauge Bosons}

\subsection{Fermions on the brane}

As discussed in the previous section, if all of the standard model
fermions are on the brane, then their couplings to the KK-gauge
bosons are enhanced by a factor of $\sqrt{2\pi kR}\sim 8.4$.  This will
lead to substantial corrections to fermion pair production through the
diagrams of Fig. 1.    In this diagram, we neglect the
n=1 weak mixing angle, which is defined as the rotation angle between 
the hypercharge and $SU(2)$ gauge bosons and their mass eigenstates. 
The reason for this is
that mixing is due to electroweak symmetry breaking, and the scale of
the KK-gauge boson masses is much, much larger.  This is similar to
the case of universal extra dimensions \cite{dobrescu} in which the
weak mixing angle for the $n=1$ states was shown to be $O(0.01)$.

\begin{figure}
\centerline{ \epsfxsize=4.0in {\epsfbox{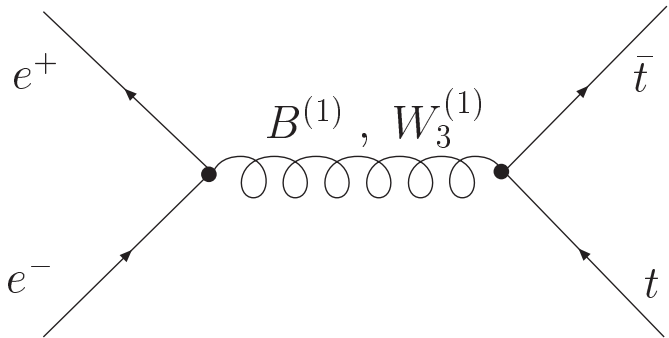}}}
    \caption{Tree-level diagrams affecting top pair production.  The 
exchanged gauge bosons are the KK-$W_{3}$ and KK-B.}
\end{figure}

The corrections to the top pair production cross section 
can be easily calculated for
the exchange of the $n=1$ KK-gauge bosons.  The result is given in
Fig. 2 for $\sqrt{s}=0.5, 1.0, 1.5$ TeV.  The expected sensitivity of
the ILC is approximately one percent, and thus the ILC will be able
to probe masses up to $120$ TeV (for $\sqrt{s}=1.0$ TeV).   Note that
the interference is destructive.   The sensitivity to high mass scales
should not be surprising, since one expects the change in the cross
section to be approximately $2\times (8.4)^{2}\times {s\over
M_{KK}^{2}}$, and a one percent sensitivity for $\sqrt{s}= 1$ TeV
gives a bound on $M_{KK}$ of $120$ TeV.

\begin{figure}
    \centerline{ \epsfxsize 4.0in {\epsfbox{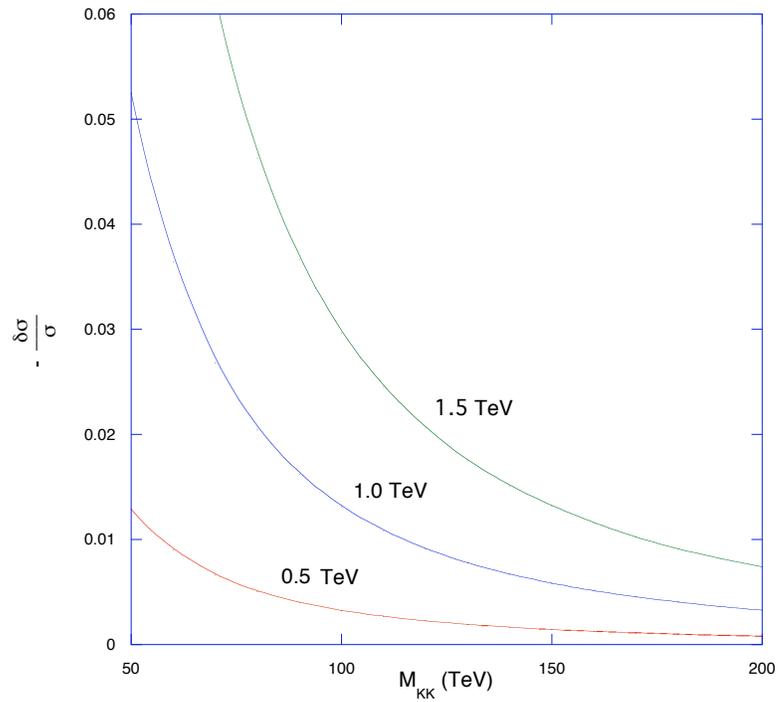}}}
    \caption{Corrections to the top pair production cross section from 
    the diagrams of Figure 1, as a function of the n = 1 KK-gauge 
    boson mass, for center of mass energies of 0.5, 1.0 and 1.5 TeV.}
\end{figure}

One can also have the $n=2, 3, \ldots$ KK-gauge bosons exchanged.  In
universal extra dimensions, the KK-gauge boson masses vary linearly
with $n$, and thus one would multiply the result by 
$\sum_{n=1}^{\infty}
{1\over n^{2}}\sim 1.6$.   In the Randall-Sundrum case, one must sum
over the zeroes of Bessel functions.   Doing this numerically, one
also gets approximately an enhancement of $1.6$.  This would increase
the bound by approximately $30$ percent, if the model isn't cut off
at higher scales.  Thus, we find sensitivity to masses up to $150$ 
TeV.

Note that there is nothing special about the top quark in this
calculation---similar results would occur for production of any
fermion pair, including muons.   Thus, one could obtain sensitivity
to even greater mass scales looking at pair production of other 
fermions. 

One could ask about the reliability of perturbation theory.  Because
of the enhancement, the effective coupling constants of the weak gauge
bosons at the TeV scale are $(8.4)^{2}\left({\alpha_{w}\over 
4\pi}\right)\sim .20$.   Depending on
coefficients, there could be significant higher order corrections.

If the fermions are not on the brane, then the electron coupling to
the KK-gauge bosons will be much weaker since the electron is further 
away from
the TeV brane.  Instead of an enhancement factor of $8.4$, the
coupling decreases \cite{gher} by a factor of roughly $5$.   This
change alone would reduce the above bound by a factor of
$\sqrt{40}$.  In addition, the top quark coupling will also be
smaller.   We consider this bound, as well as other contributions from
one-loop corrections, in the next subsection.

\subsection{Fermions off the brane}

As discussed earlier, the scenario in which fermions propagate in the 
bulk is extremely attractive, in that it provides a simple explanation 
for the fermion hierarchy.   In additon to the tree-level 
contributions of the last subsection, there are two additional 
contributions (these are also present in the on-the-wall case, but 
are substantially smaller than the tree level contributions).   One 
can calculate one-loop diagrams in which the final state top quarks exchange
KK-gauge bosons---these can be significant because the gauge bosons 
can be gluons.    The other contribution arises from mixing between 
the zero mode and KK-gauge bosons.  We consider each in turn.
\vskip 2cm
{\noindent {\it \bf Tree-Level Contributions}}
\vskip 1cm

We first consider the same diagrams as in Fig. 1.  As noted in the
previous paragraph, one expects the bound to be lowered from the
on-the-brane case by a factor of at least $\sqrt{40}$, which gives a
reach of approximately $25$ TeV.  This will be lowered further since
the top quark is not on-the-brane, and so its coupling will be
weakened.

In general, the left and right handed top quarks will have different
5-d mass terms, $c_{L}$ and $c_{R}$.   This will lead, from Eq. 
\ref{g(n) coupling}, to
different enhancements for the different chiralities.   If the 
enhancement of the left handed top quark couplings is $\alpha_{L}$, 
and that of the right handed top quark couplings is $\alpha_{R}$, one 
can then determine the cross sections and asymmetries.  

Using the 
notation of Ref. \cite{godfrey} for exchange of a neutral heavy gauge 
boson $Z^{\prime}$, the differential cross section can be 
written as
\begin{equation}
	\frac{d\sigma_L}{d\cos\theta} = \frac{\pi\alpha^2}{4s} \left\{|C_{LL}|^2(1+\cos\theta)^2 + |C_{LR}|^2(1-\cos\theta)^2\right\}
\end{equation}
where
\begin{equation}
	C_{ij} = -Q_f + \frac{C_i^e C_j^t}{c_w^2 s_w^2} \frac{s}{(s-M_Z^2)+i\Gamma_Z M_Z} + 
	\frac{\left(g_{Z^\prime}/g_{Z^0}\right)^2 C_i^{e^\prime} C_j^{t^\prime}}{c_w^2 s_w^2} 
	\frac{s}{\left(s-M_{Z^\prime}^2\right)+i\Gamma_{Z^\prime}M_{Z^\prime}}.
\end{equation}
Here, $C_{i}^{t}$ are the SM $Z^{0}$ couplings and  
$C_{i}^{t^{\prime}}$ are the $Z^{\prime}$ couplings to the top quark.   For right-handed 
electrons, one substitutes $C_{LL}\rightarrow C_{RR}$ and 
$C_{LR}\rightarrow C_{RL}$.   From this, one finds the unpolarized 
total cross section is given by
\begin{equation}
	\sigma = {\pi\alpha^2\over 3s}\left[|C_{LL}|^2+|C_{RL}|^2+|C_{LR}|^2+
	|C_{RR}|^2\right],
\end{equation}
the forward-backward asymmetry is given by
\begin{equation}
	A_{FB} = \frac{\left[\int_0^1-\int_{-1}^0\right]d\cos\theta \frac{d\sigma}{d\cos\theta}}
	{\left[\int_0^1+\int_{-1}^0\right]d\cos\theta \frac{d\sigma}
	{d\cos\theta}},
\end{equation}
and the left-right asymmetry is
\begin{equation}
	A_{LR}^f = \frac{\sigma\left(e_L^-\right)-\sigma\left(e_R^-\right)}
	{\sigma\left(e_L^-\right)+\sigma\left(e_R^-\right)}.
\end{equation}

Using these results, we 
find that the corrections to the cross section, forward-backward asymmetry and left-right 
asymmetry (using the expected value \cite{gher} of $-0.2$ for the 
change in the electron coupling to the KK gauge bosons) are given by
\begin{eqnarray}
{\delta\sigma\over\sigma}&=& (0.24\alpha_{L}+0.14\alpha_{R}){s\over 
M_{KK}^{2}}\cr\cr
\delta A_{FB}&=& (-0.04\alpha_{L} - 0.03\alpha_{R}){s\over 
M_{KK}^{2}}\cr\cr
\delta A_{LR}&=& (0.26\alpha_{L} - 0.19\alpha_{R}){s\over M_{KK}^{2}}
\end{eqnarray}

The result is plotted in Figure 3 as a function of $c_{L}$ and
$c_{R}$.   Here, we choose $M_{KK} = 10$ TeV, the results in all cases 
scale like the inverse-square of $M_{KK}$.   These results are for 
the $n=1$ KK gauge bosons.  Including the sum of all KK-modes results 
in a small change of less than $20$ percent (this is less than the 
sixty percent correction in the last subsection since for some values 
of the mass term, the couplings of higher modes can be negative).

\begin{figure}
\leftline{ \epsfxsize 2.8in {\epsfbox{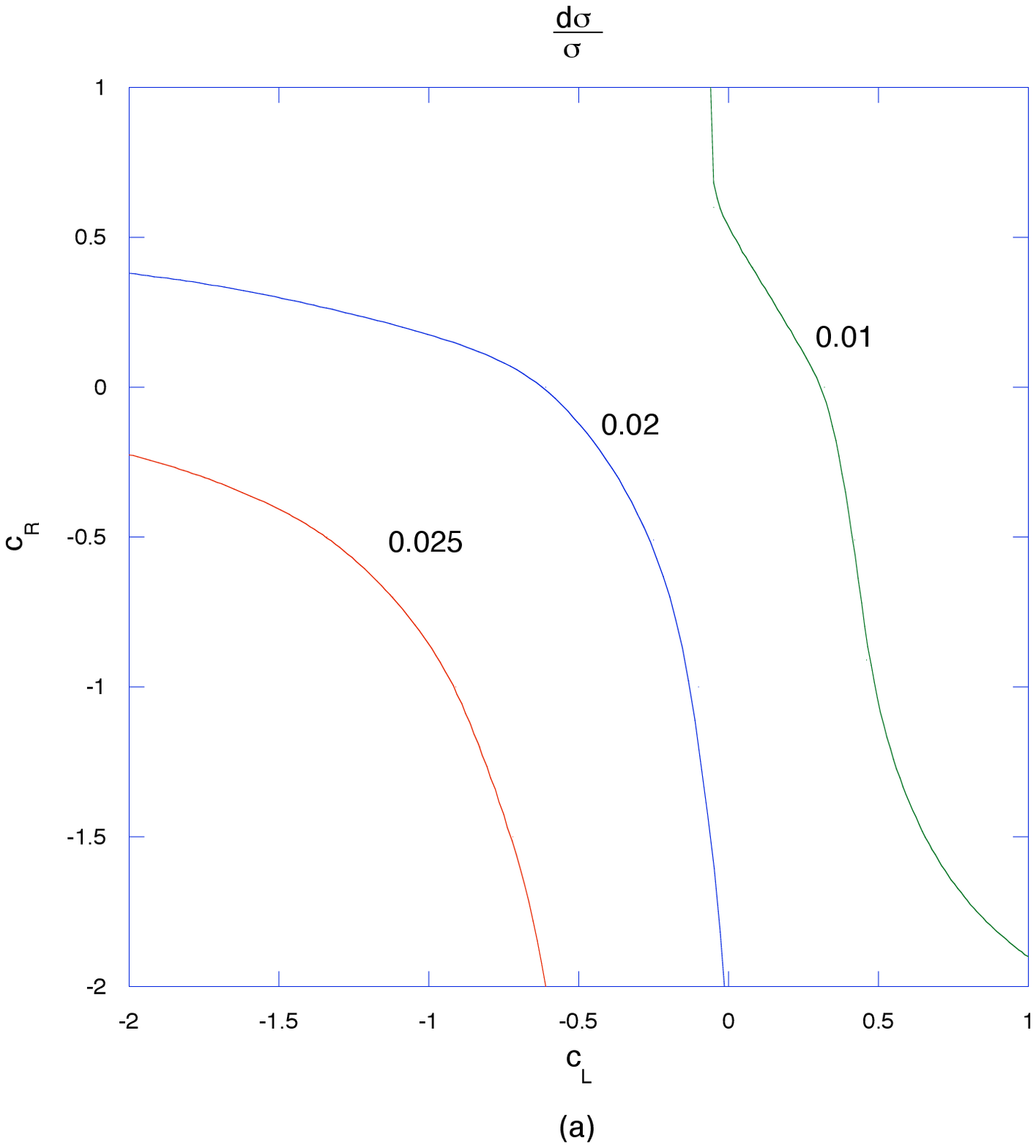}}} 
\vskip -3.2in \rightline{ \epsfxsize 2.8in {\epsfbox{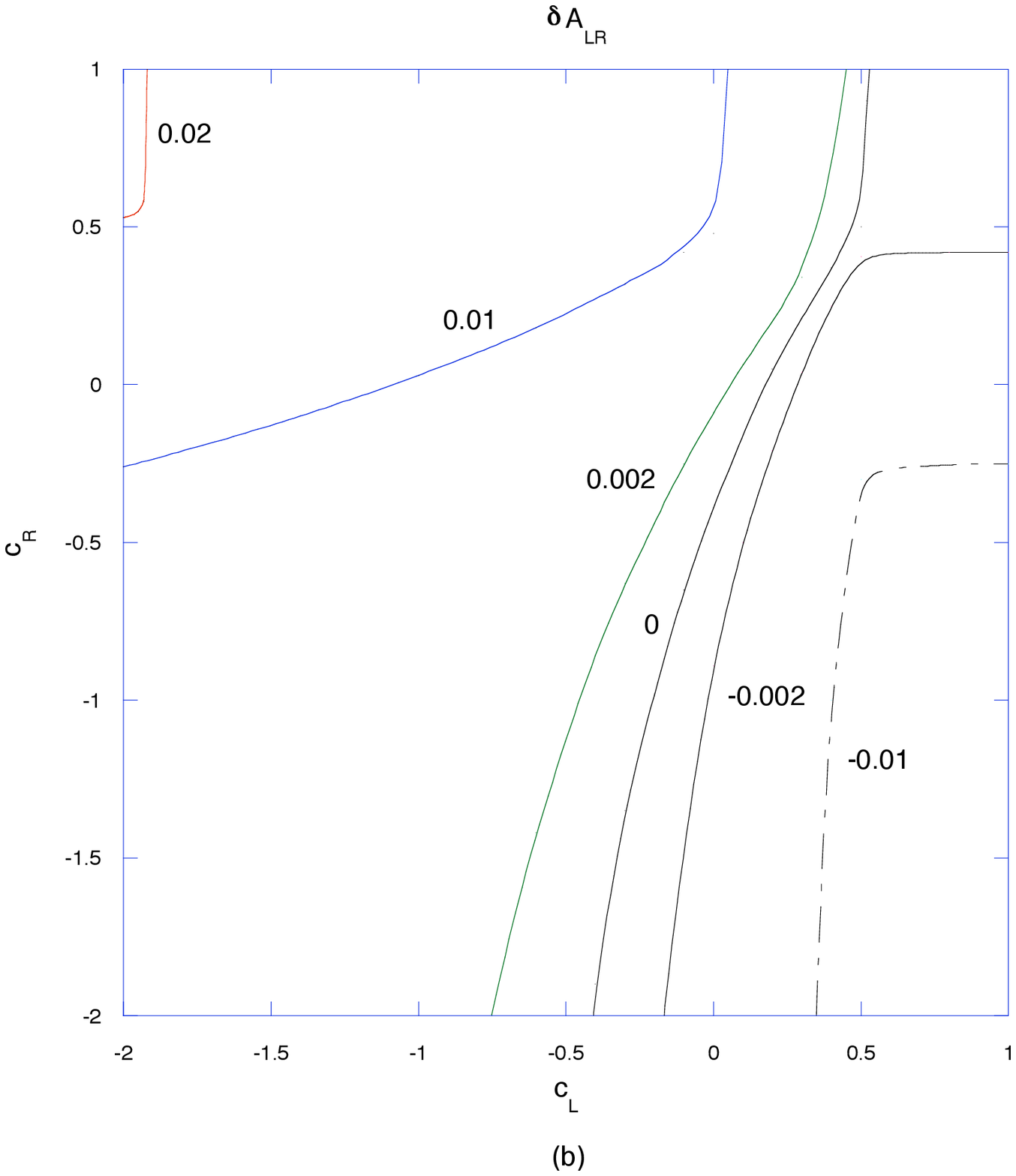}}}
   \caption{For different values of $c_{L}$ and $c_{R}$, corrections 
   to the (a) production cross-section and (b) the left-right asymmetry. 
  The  contribution to the forward-backward asymmetry is negligible.   We 
  have assumed that 
   $M_{KK}=10$ TeV; the results will scale as $1/M_{KK}^{2}$. }
\end{figure}

Depending on how 
precisely the luminosity at an ILC can be determined, a one-percent 
measurement of the cross-section is possible, and thus a reach of $10$ 
TeV for much of parameter space can be obtained (and a reach of $15$ TeV 
for some of parameter space is possible).   The forward-backward 
asymmetry is too small to be measurable.   The left-right asymmetry is 
interesting.  With a million top pairs expected in several years 
running, half from left-handed and half from right-handed electrons, 
assuming $80\%$ polarization, one could reach a sensitivity of 
approximately $0.002$ for $A_{LR}$, which would also cover
most of parameter space, for a $10$ TeV KK gauge boson mass, and would 
cover some of the space even for a $30$ TeV mass.   It should be noted 
that the ``preferred'' range of $c_{L},c_{R}$, since the right handed 
top can be much closer to the TeV brane, is for negative (or near 
zero) $c_{R}$ and 
for $c_{L}$ positive (but less than $0.5$).  A clear signature of the 
model, which could distinguish it from extra-Z models, is the absence 
of a substantial change in the forward-backward asymmetry.

These bounds could perhaps be improved substantially by including 
the effects of positron beam polarization and of top quark 
polarization \cite{godfrey}, which can increase the bounds by up to a 
factor of two.   This improvement, of course, depends on the 
design of the ILC.
\vskip 1cm
{\noindent {\it \bf One-loop Contributions}
\vskip 1cm
We now turn to one-loop corrections to the $\gtt$ and the
$\ztt$ vertices.  We start with the diagrams in Fig. 4.  The
exchanged KK-gauge boson can be either a KK-gluon, KK-$W_{3}$, or a
KK-B.  Of course, one expects the KK-gluon to have the biggest effect;
this is the KK-version of the well-known ${\alpha\over\pi}$ correction
to the value of R in hadron production.    In fact, we find this to be 
the case, but present the results for all of the diagrams for 
completeness.

\begin{figure}
\centerline{ \epsfxsize 4.0in {\epsfbox{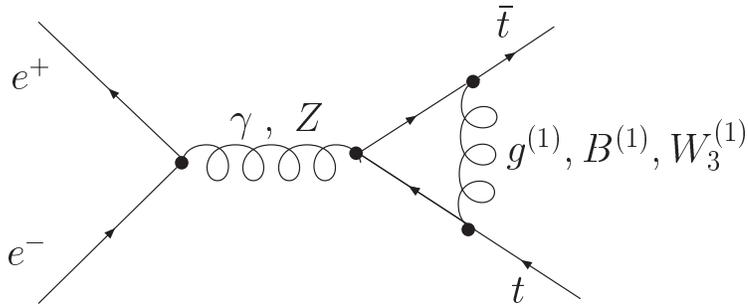}}}
    \caption{The dominant one-loop diagrams affecting top pair production.  The 
    exchanged gauge boson is either a KK-gluon, KK-$W_{3}$ or a 
    KK-B.   Corrections to the electron vertex are negligible since 
    the electron couplings to the KK-gauge bosons are suppressed.   
    Other diagrams, noted in Appendix A, do not involve KK-gluons and 
    are numerically small.}
\end{figure}

The most general interactions of the top quark with the $\gamma$ and $Z$, assuming
massless initial fermions and ignoring the (small) CP-violation, is
\begin{equation}
\Gamma_{\mu}^{V}(q^{2})=-ie\left[\gamma_{\mu}(F^{V}_{1V}(q^{2})+\gamma_{5}F_{1A}^{V}(q^{2}))+{i 
\sigma_{\mu\nu}q^{\nu}\over 2m}(iF^{V}_{2V}(q^{2}))\right]
    \label{general interactions}
\end{equation}
where $V=\gamma,Z$. As calculated in Ref. \cite{snowmass} and discussed 
by Baur \cite{baur}, these coefficients can all be bounded
at roughly the one percent level.  Baur gives the precise bounds that
can be obtained at the ILC.    However, the bounds that he lists are 
from early studies \cite{teslatdr}, where the integrated luminosity is 
either $100$ 
or $200$ ${\rm fb}^{-1}$.   We are assuming that many years of 
running at an ILC can yield an integrated luminosity of an inverse 
attobarn, and thus one can (in the extremely optimistic case of 
assuming statistical uncertainties only) scale the 
results by the square-root of the integrated
luminosity ratio for interference diagrams, and the fourth-root 
for direct terms.   Positron polarization ($50\%$) also lowers the 
limits by $25\%$, 
and a center-of-mass energy of $1$ TeV also lowers them by a factor 
of $1.5$ \cite{teslatdr}, compared to the earlier studies which assumed 
half the center-of-mass energy and no polarization.
Including these latter two effects, we take the 
range of the bounds on the coefficients to be between the values cited 
by Baur and the optimistic range given with an inverse attobarn 
luminosity.   The ranges of interest are then
\begin{eqnarray}
    F_{1V}^{\gamma} &:& .010 - .024\cr
    F_{1A}^{\gamma} &:& .003 - .006\cr
    F_{2V}^{\gamma} &:& .010 - .019\cr
    F_{1V}^{Z} &:& .003 - .006\cr
    F_{1A}^{Z} &:& .002 - .006\cr
    F_{2V}^{Z} &:& .002 - .006
    \end{eqnarray}

In principle, one could add the effects of these diagrams to the 
tree-level contribution, and calculate the resulting cross sections 
and polarization asymmetries in a unified manner.  One could 
calculate the corrections to the cross section and 
asymmetries for a given $F$; for example, one can show that the 
contribution of $F_{1V}^{Z}$ to $\delta\sigma/\sigma$ is negligible, 
whereas the contribution of $F_{1A}^{Z}$ is roughly 
$\delta\sigma/\sigma = 2.2\ \delta F_{1A}^{Z}$.   However, the 
tree-level contribution is similar to that of an extra Z boson for 
which virtually all studies generally refer to cross sections and asymmetries, while 
the one-loop contribution involves anomalous $\gamma$ and $Z$ 
interactions, for which studies generally refer to the above form 
factors.   Furthermore, the sensitivity to changes in the cross 
section and asymmetries were calculated using different assumptions 
about the collider than those for the sensitivity to changes in the form factors.
Since the detailed specifications of the ILC and its 
detectors are not yet known, we are simply referring to previous 
studies and thus keep the contributions separate.    A more detailed 
unified study, including top quark and positron polarization asymmetries 
would be valuable and could make our results more precise.

\begin{figure}
\vskip -1.0cm
\leftline{ \epsfxsize 2.6in {\epsfbox{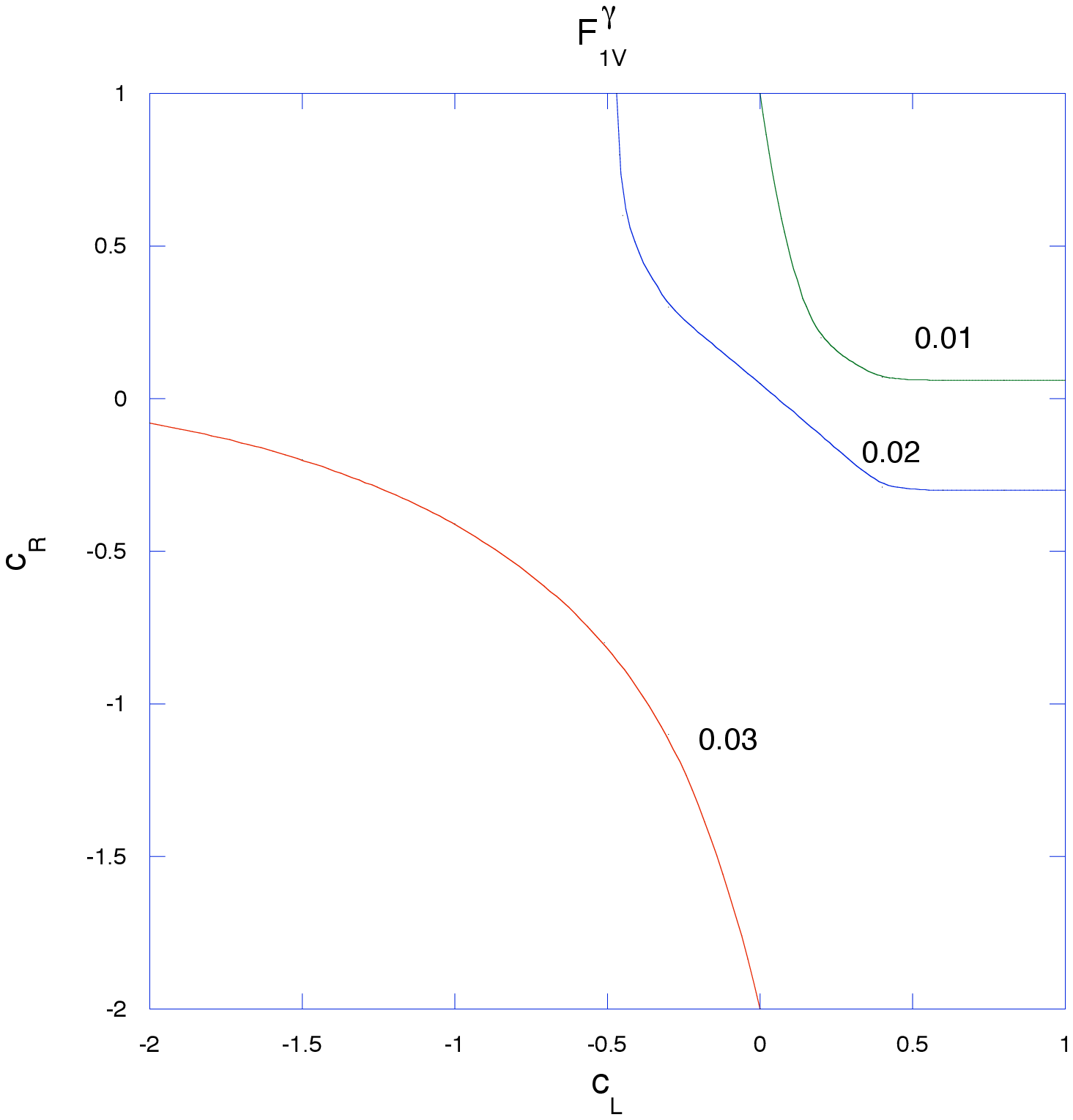}}} 
\vskip -2.7in \rightline{ \epsfxsize 2.6in {\epsfbox{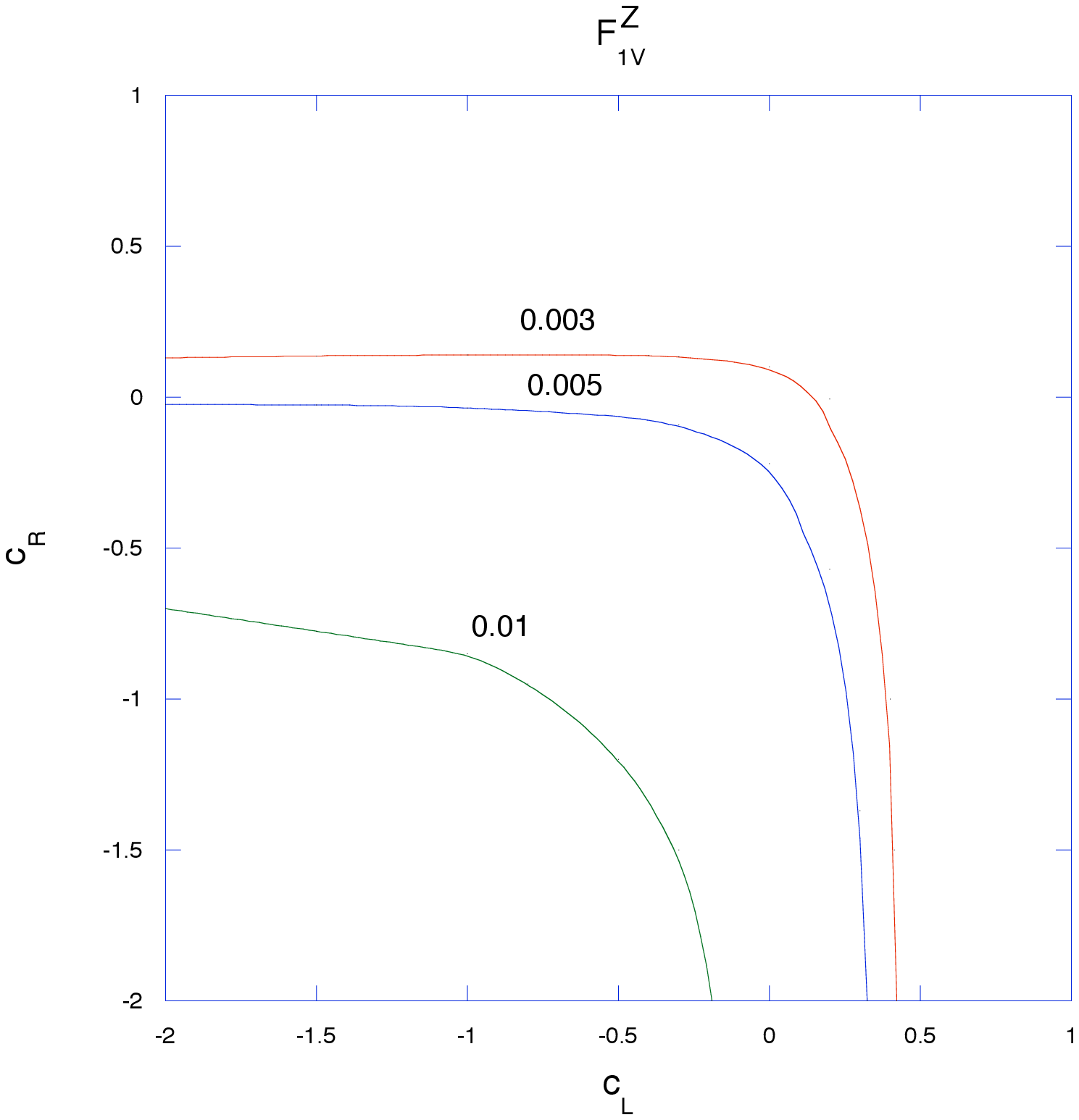}}}
\vskip 0.4cm
\leftline{ \epsfxsize 2.6in {\epsfbox{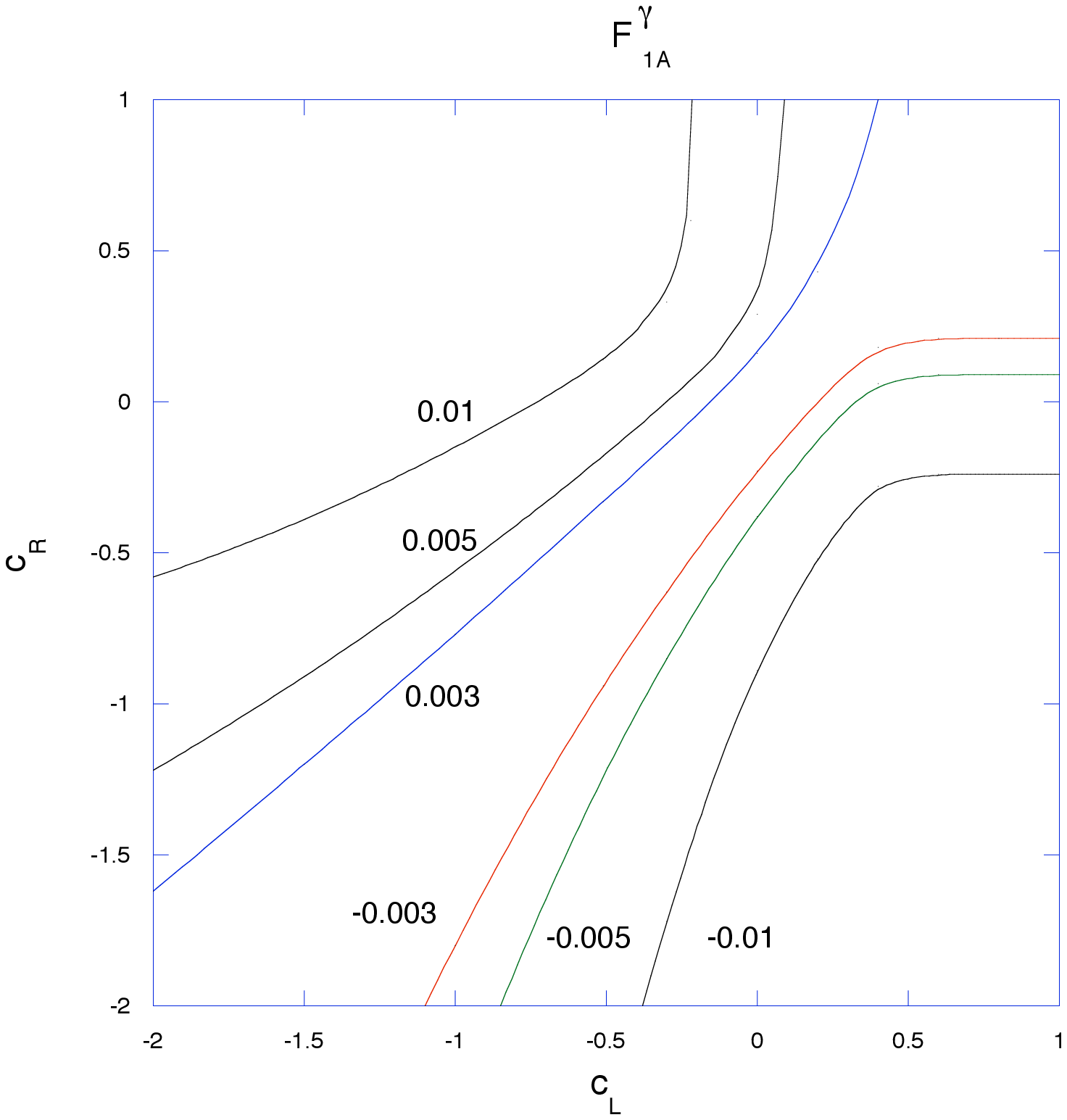}}} 
\vskip -2.7in \rightline{ \epsfxsize 2.6in {\epsfbox{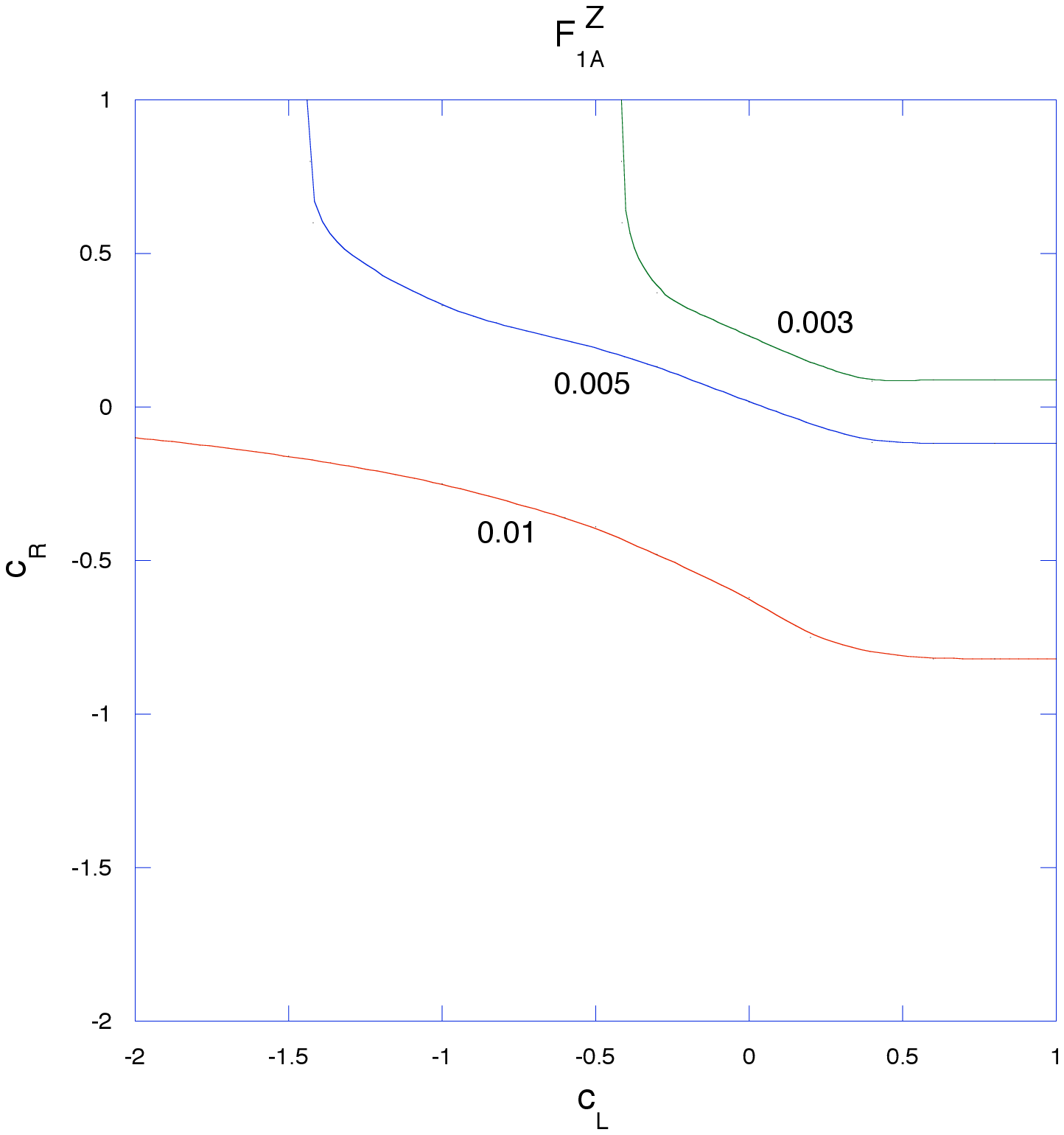}}}
\vskip 0.4cm
\leftline{ \epsfxsize 2.6in {\epsfbox{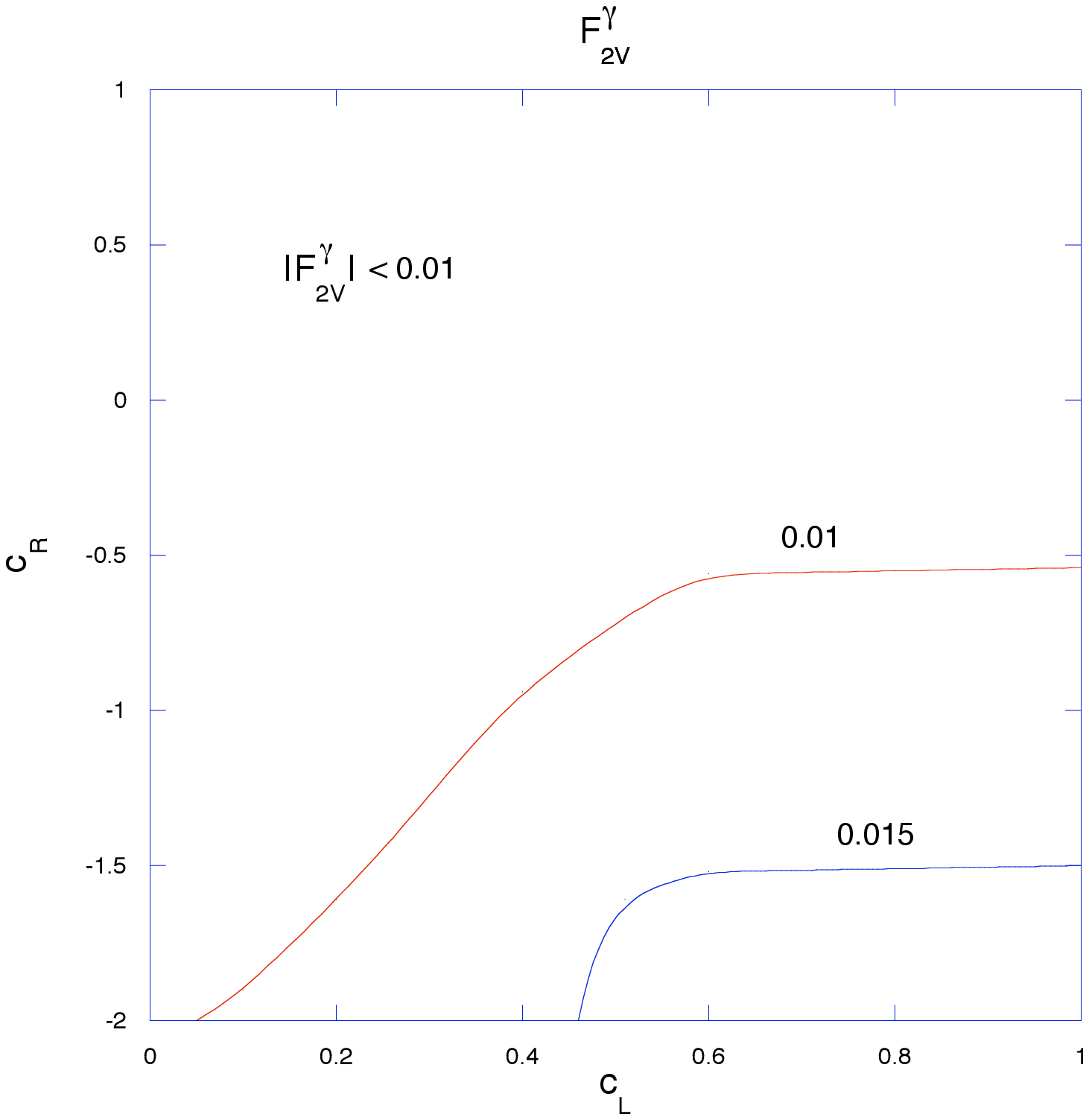}}} 
\vskip -2.7in \rightline{ \epsfxsize 2.6in {\epsfbox{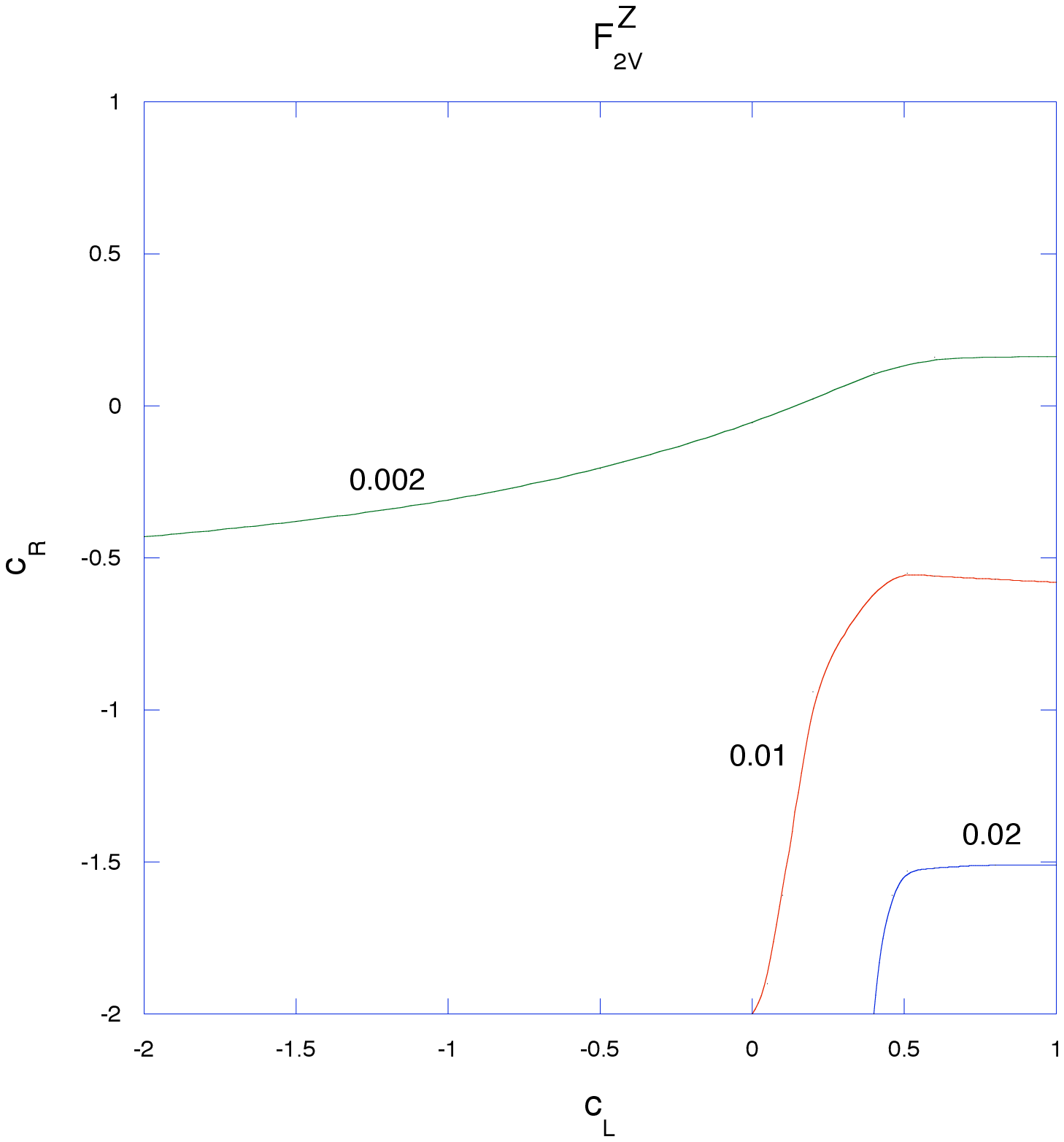}}}
    \caption{Contributions to the $\gamma$ and Z form factors as a 
    function of $c_{L}$ and $c_{R}$, for $M_{KK}=5$ TeV.}
\end{figure}

The detailed calculations are given in Appendix A.
For a given value of $c_{L}$ and $c_{R}$, we can find the enhancements
of the couplings of the left and right-handed top quarks, determine
the value of $C$ and $\alpha$ in the vertex (see Appendix A), plug into the
expressions and determine the effect on the six parameters in Eq.
\ref{general interactions}, for $q^{2}=s=1\ {\rm  TeV}^{2}$.  
As in the tree-level case, including higher order terms will increase 
the mass reach by approximately $20\%$---more precision is 
unnecessary since higher order corrections (such as double KK-gluon 
exchange) will likely  have a bigger effect.   The results are plotted in 
Figure 5,  assuming $M_{KK}=5$ TeV.  We see that the most 
sensitive coefficients are the couplings of the $Z$, for which sensitivities to 
$M_{KK}=5$  TeV are reached for most of parameter space.   However, 
we have found that for $M_{KK}=10$ TeV, only a small sliver of parameter 
space is sensitive. 
These results are substantially weaker than the results for the tree-level 
contribution of the last subsection.

\vskip 1cm
{\noindent {\it \bf Contributions from mixing}}
\vskip 1cm
The most detailed discussion of top pair production at a 
linear collider in the Randall-Sundrum model was by Agashe, Delgado, 
May and Sundrum (ADMS) \cite{agashe1}, which was recently summarized by 
Agashe \cite{agashe2}.   They discuss the contributions from mixing 
between the Z-boson and the KK-Z bosons.  This mixing occurs from the 
Higgs vev.   The biggest effect is on the right-handed top quark 
coupling, and they find that
\begin{equation}
    {\delta(g_{Z}^{t_{R}})\over g_{Z}^{t_{R}}} \sim {m_{Z}^{2}\over 
    (0.41M_{KK})^{2}}{1-2c_{R}\over 3-2c_{R}}\left( {-k\pi R\over 
    2}+{5-2c_{R}\over 4(3-2c_{R})}\right)
    \end{equation}
    It is straightforward to convert this into a shift in $F_{1V}^{Z}$ 
and $F_{1A}^{Z}$, 
\begin{equation}
    F_{1V}^{Z}=F_{1A}^{Z}= -{\tan\theta_{W}\over 3} {\delta(g_{Z}^{t_{R}})\over g_{Z}^{t_{R}}} 
    \label{fvfa}\end{equation}

For a KK-gauge boson mass of 5 TeV, this gives a result for 
$F_{1V}^{Z}$ and $F_{1A}^{Z}$  which ranges from 0 at $c_{R}=1/2$, to $0.002$ at 
$c_{R}=0$, to $0.004$ at $c_{R}=-0.2$.  We see that the $5$ TeV mass 
scale can barely be reached for the $c_{R}<0$ part of parameter space, and 
thus could have a greater reach than the one-loop contributions for 
some of the parameter space.  But it is substantially weaker than the 
tree-level contribution.   As we will see in the next section, 
however, the effects of mixing between the top quark and the KK-top can be 
substantially larger, and could be competitive with the tree-level 
contribution.

\section{Effects of KK Fermions and Brane Kinetic Terms}

In our analysis, we have only included the effects of KK-gauge bosons.
As noted in Section 2, the masses of the KK-gauge bosons are related 
to the zeroes of Bessel functions of order $1$, while the masses of 
the KK-fermions are related to zeroes of Bessel functions of order 
$|c\pm 1/2|$.  In the absence of brane kinetic terms, the masses of 
the KK-tops are thus related to those of KK-gauge bosons, and their 
effects must be considered.  In particular, the masses of the 
left-handed KK-tops \cite{agashe1,agashe2} are given by $m_{t^{n}_{L}}\sim \pi 
k e^{-k\pi r_{c}}(n-c_{L}/2)\sim 1.28 m_{KK} (n-c_{L}/2)$, where 
$m_{KK}$ is the $n=1$ KK-gauge boson mass\footnote{In Ref. 
\cite{agashe2}, there are two typographical errors in Eq. 16--the 
factor of $\sqrt{1/2-c_{L}}$ should be in the denominator and the 
factor of $0.78$ should be $1/0.78$.  There are purely typographical 
and do not affect the results.}.   For $n=1$ and $c_{L}=0.4$, this 
gives virtually equal $n=1$ KK-top and KK-gauge boson masses.

Clearly, the results from tree-level KK-gauge boson exchange will not 
be affected, except for small mixing effects, by KK-top 
contributions.   There will, however, be contributions to the 
one-loop diagrams of Figure 4, in which the internal top quark lines 
are replaced by KK-top quark lines.   We have calculated the effects 
of these contributions, and find them to be smaller, in all cases, 
than the previous results.

A much bigger effect arises from mixing between the top quark and the 
KK-top quark.  This arises from mixing of the zero-mode $t_{R}$ with 
the KK-$t_{L}$ through the Higgs vev, and is discussed in detail by 
Agashe \cite{agashe2}.
Using Eq. \ref{fvfa}, Agashe's result can be written as
\begin{equation}
    \delta F_{1V}^{Z}=F_{1A}^{Z}\sim \sum_{n}{-1\over 2\sin 
    2\theta_{W}}\left( {m_{t}\over 
    m_{t_{L}^{(n)}}}\right)^{2}\left( {1-e^{-2k\pi R(1/2-c_{L})}\over 
    1/2 - c_{L}}\right).
    \end{equation}
This is plotted as a function of $c_{L}$ for several masses in Figure 
6, where the sum over the KK-modes has been included.  The range 
$c_{L}>0.5$ is exceeedingly disfavored, since the Yukawa coupling of 
the top quark would then be exponentially suppressed.    We see that for 
$c_{L}=0.4$, a reach of $10$ TeV is barely possible, with the
optimistic assumptions discussed earlier for the reach of the ILC.  
For $c_{L}$ very close to $0.5$, however, the reach can exceed that of 
the tree-level KK-gauge boson exchange.

\begin{figure}
\centerline{ \epsfxsize 4.0in {\epsfbox{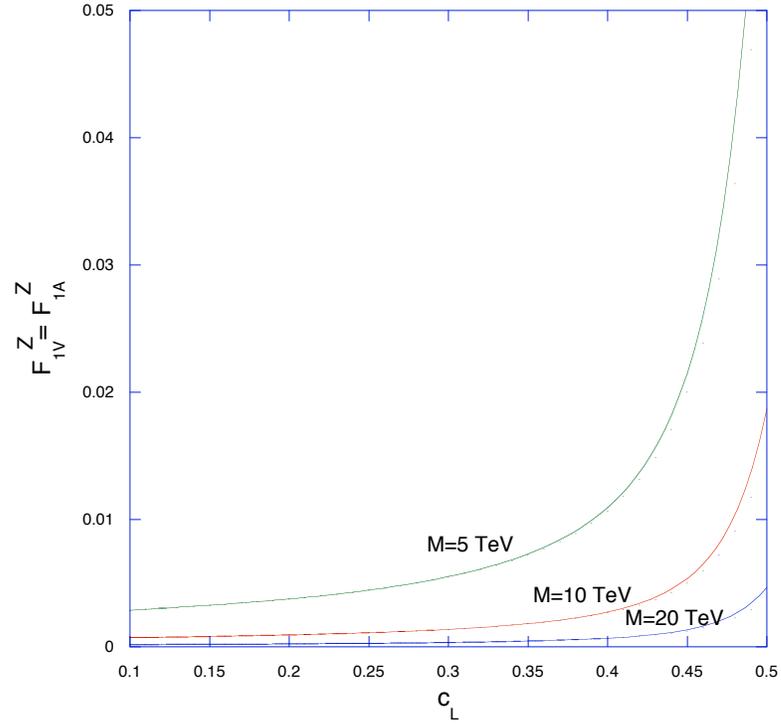}}}
\caption{Effects on the Z form factors due to top/KK-top mixing as a function of $c_{L}$ for 
various values of the KK-mass.  A high luminosity ILC should have a 
sensitivity of
$0.006$ to these form factors, and could optimistically reach $0.003$.}
\end{figure}

Thus, mixing can give a reach which can be larger than that of the tree-level 
KK-gauge boson exchange, but only in the upper end of the  $0.4\leq 
c_{L}\leq 0.5$ range.  Although this seems narrow, it is a particularly 
interesting range of $c_{L}$. If $c_{L}$ were larger, the Yukawa 
coupling would be suppressed and the top mass would be too small, and 
if it were much smaller, there might be dangerous contributions to the 
$\overline{b}bZ$ vertex.   A word of caution is that the large 
mixing can cause problems with precision electroweak fits, although a 
custodial SU(2) symmetry or brane kinetic terms can ameliorate the 
problems (if there is a custodial SU(2) symmetry, one should include 
effects of the $Z^{\prime}$ as well).   Mixing contributions between the 
zero-mode $t_{L}$ and the KK-$t_{R}$ are expected to be small since 
$c_{R}$ is not expected to be in this range.   Note that a clear 
signature of the dominance of mixing would be the equality of 
the contributions to $F_{1V}^{Z}$ and $F_{1A}^{Z}$.  Here, one looks 
for deviations in the right-handed top quark couplings, and this 
might require determination of the top quark polarization.  Previous 
analyses have looked at $F_{1V}^{Z}$ and $F_{1A}^{Z}$ separately 
(assuming one is nonzero and all others vanish)---here a more unified 
analysis for the ILC would be welcomed.

Finally, we consider the effects of brane kinetic terms (BKTs).  A 
detailed discussion of these terms in flat space can be found in Ref. 
\cite{delaguila}.    In the context of Randall-Sundrum models, two 
papers by Carena, Delgado, Ponton, Tait and Wagner (CDPTW) \cite{care,care2} 
have extensively studied BKTs and their effects on 
phenomenology.   The BKTs for fermions 
arise in the 5D action
\begin{equation}
    S=-\int d^{4}x \int_{0}^{\pi 
    R}dy\sqrt{-G}\left({i\bar{\Psi}\Gamma^{A}e^{M}_{A} D_{M}\Psi + 
    im(y)\bar{\Psi}\Psi + 2\alpha_{f}\delta(y-\pi R)}\bar{\Psi}_{L} 
    \gamma^{a}e^{\mu}_{a}\partial_{\mu}\Psi_{L}\right)
    \end{equation}
 where $\Gamma$ and $\gamma$ are the 5D and 4D Dirac matrices, and the 
 last term is the BKT.  Here, the $\delta$ function is normalized so 
 that $\int_{0}^{\pi R}2\delta(y)dy=1$.   The coefficient, 
 $\alpha_{f}$, has dimensions of length.   Note that this is an IR-brane BKT, 
 whereas a UV brane BKT would be proportional to $\delta(y)$, but one 
 expects the UV brane BKTs to be less phenomenologically relevant.  
 More details can be found in CDPTW.   

 One can also have gauge field BKTs.   For a $U(1)$ gauge group, the 
 relevant part of the action is 
 \begin{equation}
     S=-\int d^{4}x\int_{0}^{\pi R} dy {1\over 2} B_{\mu}{\cal 
     O}^{\mu\nu}B_{\nu}
     \end{equation}
     where
     \begin{equation}
	 {\cal O}^{\mu\nu}={1\over g_{5}^{2}}\left( P^{\mu\nu} + 
     \eta^{\mu\nu}\partial_{y}(e^{-2\sigma}\partial_{y}) + 
     2\delta(y)r_{UV}P^{\mu\nu} + 2\delta(y-\pi 
     R)r_{IR}P^{\mu\nu}\right)
     \end{equation}
     and $P^{\mu\nu}\equiv 
     \eta^{\mu\nu}\partial^{2}-\partial^{\mu}\partial^{\nu}$.   Note 
     that we have explicitly included both UV and IR BKTs.
     
 CDPTW \cite{care,care2} use these actions and find all of the 
 KK-masses, wavefunctions and couplings in the model, and the reader 
is referred to those papers for the full expressions.   They find that 
the IR BKTs repel the KK-wavefunctions from the IR brane, thus 
reducing the couplings of the zero-mode fermions to the KK gauge 
bosons.  As a result, the effects on precision tests is reduced, and 
KK-masses of the order of a few TeV (and thus in reach of the 
LHC) become allowed.  In addition, BKTs can also make the model more 
compatible with grand unification.   Relatively large BKTs (of 
order $\pi R$) are needed to have a substantial impact, but such 
terms are not unnatural.

As discussed in the introduction, our approach in this paper is to 
consider KK-masses which are out of reach of the LHC.   The effect of 
the BKTs discussed by CDPTW is then to reduce the coupling of fermons 
to KK-gauge bosons, and 
thus lower the effects in top pair production.  In short, we have added some 
parameters to the model which, if large enough, can substantially 
weaken our bounds .   

One interesting feature concerns the conformal limit 
($c_{L}=c_{R}=1/2$).  At this point, without BKTs, the coupling of the 
zero mode fermions to the KK-gauge bosons vanish, and all of the 
contributions we discussed (involving KK-gauge bosons) vanish (as well 
as many contributions to electroweak precision tests).  This 
is because the fermion zero-mode wavefuntion is flat, and thus 
proportional to the gauge zero-mode wavefunction, which is orthogonal 
to the KK-gauge boson wavefunctions.   This was first noticed in the 
Randall-Sundrum model in Ref. \cite{agashe1}, and for Higgsless 
models in Ref. \cite{higgsless}.   With 
BKT's however, unless the gauge and fermion BKTs are identical, the 
fermion and gauge boson orthogonality conditions will differ, and the 
couplings won't vanish in the conformal limit.   Whether the 
couplings are large enough to make a measureable contribution 
depends, of course, on the size of the BKTs.

\section{Discussion and Conclusions}

The Randall-Sundrum model is one of the most promising approaches to 
solving the gauge hierarchy problem.   The five-dimensional spacetime 
compactified on an orbifold, with a slice of ${\rm ADS}_{5}$ 
describing the bulk geometry, can not only explain a large hierarchy 
but also may naturally arise from string theory.  
The original form of the model had 
all of the Standard Model particles on the TeV brane, but there has 
been much interest in versions of the model in which gauge bosons 
and/or fermions can propagate.  Such models can also naturally explain 
the fermion mass hierarchy.   In this case, the KK excitations of 
the gauge bosons and/or fermions can have significant 
phenomenological consequences.

Most analyses of the phenomenology of the Randall-Sundrum models have 
looked at the effects of the KK excitations on precision electroweak 
constraints, and there have been many interesting modifications to the 
model which ameliorate many of these constraints.  This can allow the 
KK excitations to be within reach of the LHC.  The most appealing of 
these modifications include imposing a custodial $SU(2)$ gauge symmetry in 
the bulk (which may come from a global $SU(2)$ symmetry in the 
AdS/CFT related conformal theory), or by adding gauge or fermion 
brane kinetic terms, or both.

Our approach is different.  We will suppose that the KK excitations 
have masses well in excess of 5 TeV, and are thus out of range of the 
LHC.  We also do not concern ourselves with precision electroweak 
constraints (which may still be signficant in the $5-15$ TeV mass 
range), assuming that one of the modifications discussed above can 
ameliorate the constraints if necessary.   We have argued that top 
pair production could be the first signature of these excitations, 
since the top quark, due to its large mass, must be close to the TeV 
brane and thus will feel the effects of these excitations more 
strongly than other fermions.

We have calculated top pair production at the ILC in the Randall-Sundrum model. 
Note that in many versions of the model, such as the version with a 
custodial $SU(2)$ symmetry or versions with extended gauge or fermion sectors,
there will be additional fields which 
could affect top pair production.   Unless there is destructive 
interference plus some tuning, however, such fields are likely to 
increase the bounds.  For simplicity, we have only considered the 
KK-excitations of standard model particles.

When all fermions are on the TeV brane, direct KK-gauge boson exchange 
gives a sensitivity to KK-gauge boson masses up to 150 TeV.   The most 
attractive models, though, are those in which fermions propagate in 
the bulk.   In this case, the tree-level KK-gauge boson exchange 
diagram still dominates for much of parameter-space, but the reach is 
much smaller, since the electron coupling is much weaker.  We found the change in the cross-section and 
left-right asymmetry as a function of the fermion mass paramters and 
the KK-gauge boson mass, and obtained a sensitivity to KK-gauge boson 
masses of approximtely 10-20 TeV, depending on the mass parameters.

We then considered the one-loop diagrams in which KK-gauge bosons are 
exchanged by the top quarks in the final state.   The dominant 
diagram is due to KK-gluon exchange.   These will affect the $\gamma$ 
and $Z$ form factors, and we find sensitivity in much of 
parameter-space to $5$ TeV KK-gauge boson masses, but $10$ TeV masses 
are out of reach.    The effects of KK-fermions on these results is 
small.

Finally, mixing between the top quarks and the KK-tops can be 
substantial in the narrow window in which $c_{L}$ is between 0.3 and 
0.5.  Although this window is narrow, it is in the phenomenologically 
preferred range. The reach can exceed $10$ TeV for some of this range.

A more detailed phenomenological analysis is needed.  Effects of positron polarization and top quark 
polarization have not really been included, the experimental 
sensitivities to the various form factors were determined by assuming 
that only one was nonzero, the relationship between those form 
factors and experimentally observed quantities is unclear (in view of 
different assumptions made).  The basic version of the Randall-Sundrum 
model has only three parameters--$c_{L}, c_{R}$ and $M_{KK}$, with 
 brane kinetic terms playing a role if they are sufficiently 
large.   This is a sufficiently small parameter set that an event 
generator could be constructed.  Recently, a version of Pythia for 
Universal Extra Dimensions \cite{pythia} was developed; such a tool 
could be developed for this model.   Certainly, one expects models 
with Kaluza-Klein excitations to behave in some sense like extra-Z 
models (as in tree-level exchange), and in some sense like anomalous 
gauge boson couplings (as in the one-loop diagrams and in mixing), so 
a Pythia-type gnerator would be helpful.

We are very grateful to Kaustubh Agashe, Chris Carone, Csaba Csaki, Josh 
Erlich, and Frank Petriello for extensive discussions.  This work was 
initiated at the Aspen Center for Physics and was supported by the 
National Science Foundation grant PHY-023400.

\newpage
 \renewcommand{\theequation}{A-\arabic{equation}}
  \setcounter{equation}{0}  
  \section*{APPENDIX}  

The diagrams in Fig. 4 are
calculated.  The counterterms will be determined by the requirement
that $\Gamma_{\mu}^{\gamma}(q^{2}=0)=-{2\over 3}ie_{R}\gamma_{\mu}$ 
and
$\Gamma_{\mu}^{Z}(q^{2}=M^{2}_{Z})=-{g_{R}\over 
4\cos\theta_{WR}}\gamma_{\mu}(1-{8\over 
3}\sin^{2}\theta_{WR}+\gamma_{5})$.

We let the coupling of the gauge boson to the top quark be
$C \gamma_{\mu}(1-\alpha\gamma_{5})$.  Note that the fact that the 
chiralities may
have different enhancements implies that even the KK-gluon will not
necessarily couple in a vector-like manner.   The numerator of the 
massive vector
propagator does contain a $k_{\mu}k_{\nu}/M^{2}$ term, but the
divergences from this term are cancelled by the counterterms, and the 
finite parts are negligible.  The
corrections to the $\gtt$ vertex due to the diagrams in 
Fig.
4 is given by
\begin{equation}
    {iC^{2}\over 16\pi^{2}}\left({2\over 3}e\right) 
\int_{0}^{1}dx\int_{0}^{1-x}dy
    \left( T_A^\mu \left({1\over\Delta_0}-{1\over\Delta}\right) + 
{T_B^\mu\over 2} \ln{\Delta_0\over\Delta} - {q^2 
T_q^\mu\over\Delta}\right)
\end{equation}
where $\Delta_{0}\equiv M^{2}(1-x-y)+m^{2}(x+y)$,
$\Delta=-q^{2}xy+\Delta_{0}$, $M$ is the KK-gauge boson mass, $m$ is
the top quark mass and
\begin{eqnarray}
    T_{A}^{\mu} &= & 2m^{2}\left(-(x+y)^{2}(1+\alpha^{2}) + 
2(x+y)(-1+3\alpha^{2})+4(1-\alpha^{2})\right)\gamma^{\mu}\cr
    &+& 4\alpha m^{2}(x+y)(2-x-y)\gamma^{\mu}\gamma_{5}\cr
    &+& 
2m[(x+y)(x+y-1)+\alpha^{2}((x+y-1)(x+y-4))](i\sigma^{\mu\nu}q_{\nu})\cr
    T_{B}^{\mu} &= & 
4(1+\alpha^{2})\gamma_{\mu}-8\alpha\gamma^{\mu}\gamma_{5}\cr
    T_{q}^{\mu} &= & -2(xy-x-y+1)(1+\alpha^{2})\gamma^{\mu}+ 
4\alpha(xy-x-y+1)\gamma_{\mu}\gamma_{5}.
    \end{eqnarray}
The corrections for the $\ztt$ vertex are given by
\begin{equation}
    {i C^2\over 16\pi^2} \left(g\over4\cos\theta_W\right) 
\int_0^1dx\int_0^{1-x}dy 
    \left(Z_A^\mu\left({1\over\Delta_{M_Z}}-{1\over\Delta}\right) 
+Z_B^\mu\ln{\Delta_{M_Z}\over\Delta} 
+Z_q^\mu\left({M_Z^2\over\Delta_{M_Z}}-{q^2\over\Delta}\right)\right)
\end{equation}
where $\Delta_{M_Z} = -M_Z^2 x y + \Delta_0$, $M$ is the mass of the 
KK-gauge boson, and $m$ is the top mass and 
\begin{eqnarray}
    Z_A^\mu & = & 
m^2\left(4\alpha(1+2\alpha)+2(x+y)\left(A(1-3\alpha^2)+2\alpha\right) 
- (x+y)^2B_1\right)\gamma^\mu\cr
    &+& m^2\left(4\alpha A+8y(1+\alpha^2-2\alpha A)+(x+y)^2B_2\right) 
\gamma^\mu\gamma_5\cr
    &+& 
m\left(2\left(A(1+\alpha^2)-2\alpha\right)+(x+y)\left(A(1+5\alpha^2)-6\alpha\right) 
+ (x+y)^2B_1\right) i\sigma^{\mu\nu}q_\nu\cr
    Z_B^\mu & = & 2\left(A(1+\alpha^2)-2\alpha+2\alpha^2A{m^2\over 
M^2}\right) \gamma^\mu\cr
    &+& 2\left(1+\alpha^2-2\alpha A+2\alpha^2{m^2\over 
M^2}\right)\gamma^\mu\gamma_5\cr
    Z_q^\mu & = & 
\left(\left(A(1+\alpha^2)-2\alpha\right)(1-x-y)-xyB_1\right)\gamma^\mu\cr    
&+& \left((1+\alpha^2-2\alpha A)(1-x-y)-xyB_2\right)\gamma^\mu\gamma_5
\end{eqnarray}
where $A=1-{8\over3}\sin^2\theta_W$, 
$B_1=A(1+\alpha^2)-2\alpha+4\alpha^2A{m^2\over M^2}$, and 
$B_2=1+\alpha^2-2\alpha A+4\alpha^2{m^2\over M^2}$.

A diagram not shown in Fig. 4 is the vacuum polarization diagram, 
in which the photon or Z propagator goes into a top quark loop, and 
then back to a KK-B or KK-$W_{3}$.   We have calculated the 
contribution of this diagram and found it to be substantially smaller 
than the diagrams considered.

There is also the  diagram in Fig. 4 in which the internal
lines are b-quarks and the charged KK-W boson is exchanged.  This 
gives (assuming $V_{tb}=1$)
\begin{equation}
    -{1\over3}{i e\over16\pi^2}\left({g^{(1)}\over2\sqrt{2}}\right) 
\int_0^1dx\int_0^{1-x}dy 
\left(W_A^\mu\left({1\over\Delta_{M_i}}-{1\over\Delta_W}\right) 
+{W_B^\mu\over2}\ln{\Delta_{M_i}\over\Delta_W} 
+W_q^\mu\left({M_i^2\over\Delta_{M_i}}-{q^2\over\Delta_W}\right)\right)
\end{equation}
where $\Delta_W=-q^2xy-m^2(x+y)(1-x-y)+m_b^2(x+y)+M^2(1-x-y)$, 
$\Delta_{M_i}=-M_i^2xy-m^2(x+y)(1-x-y)+m_b^2(x+y)+M^2(1-x-y)$, $M$ is 
the mass of the KK-gauge boson and $m$ is the top quark mass as 
before, here $m_b$ is the mass of the bottom quark and
\begin{eqnarray}
    W_A^\mu & = & -2\left(m^2(1-a)(1-x-y)^2+m_b^2(1+a)\right) 
\gamma^\mu \cr
    &-& 2\left(m^2(1-a)(1+x+y)^2-m_b^2(1+a)\right) \gamma^\mu\gamma_5 
\cr
    &+& 2m(1-a)\left(2-3(x+y)+(x+y)^2\right) i\sigma^{\mu\nu}q_\nu \cr
    W_B^\mu & = & 4(1-a)\gamma^\mu - 4(1-a)\gamma^\mu\gamma_5\cr
    W_q^\mu & = & -2(1-a)(1-x-y+xy)\gamma^\mu + 
2(1-a)(1-x-y+xy)\gamma^\mu\gamma_5
\end{eqnarray}
For $\gtt$, $M_i=0$ and $a=0$
For the $\ztt$, $M_i=M_Z$ and $a={1\over{4\over3}\sin^2\theta_W}$.

Finally, there is one diagram that we have not discussed.  The 
$\gamma$ or $Z$ can convert into a pair of charged $KK-{W}$-bosons, 
which then exchange a b-quark and convert into a top pair.
As noted earlier, the 
diagram in Figure 4 for KK-gluon exchange completely dominates the 
results, and the finite contribution of this ``2-W'' diagram is 
negligible.  However, here the divergences are not  removed by the 
counterterms.  This should not be surprising.  We have used the gauge 
choice in which $A_{5}=0$.  This is the unitary gauge, and is 
problematic for evaluating loop diagrams, since the gauge boson 
propagtors have bad high-energy behavior (the finite S-matrix 
only results from cancellations among divergent Green's functions).  
This has not been a problem for the other 
diagrams.   In this case, one should use another gauge, such 
as the `t Hooft-Feynman gauge, and include the higher modes of the 
$A_{5}$ field.   Since the finite part of the diagram is so much 
smaller than that from the KK-gluon exchange (due to a much weaker 
coupling and two heavy fields in the loop rather than one), we will 
not include this diagram.   A nice discussion can be found in Ref. 
\cite{randallschwartz}.

\end{document}